# Effects of Disorder on the Energy Landscape and Motional Mechanisms Involved in Lithium Ion Dynamics and Transport in Solid Electrolytes: Li$_{5.5}$PS$_{4.5}$Cl$_{1.5}$ Argyrodite as a Case Study

Mohammad Ali Badragheh,[a] Vanessa Miß,[b] Bernhard Roling,[b] Michael Vogel[a,*]

[a]Institute for Condensed Matter Physics, Technische Universität Darmstadt, 64289 Darmstadt, Germany
[b]Department of Chemistry, Philipps-Universität Marburg, 35032 Marburg, Germany

[*]michael.vogel@pkm.tu-darmstadt.de

**Abstract**

$^7$Li NMR diffusometry and relaxometry are combined with electrochemical impedance spectroscopy to compare the mechanisms for the dynamics and transport of lithium ions in disordered and crystalline electrolytes with argyrodite composition Li$_{5.5}$PS$_{4.5}$Cl$_{1.5}$. The *dc* conductivity of a disordered sample prepared by ball milling amounts to 0.76 mScm$^{-1}$ at room temperature, which is substantially lower than that of two previously studied crystalline argyrodites differing in the order of the anion sublattice due to various heat treatments. However, the activation energy of the *dc* conductivity is smaller for ball-milled disordered Li$_{5.5}$PS$_{4.5}$Cl$_{1.5}$ ($E_{dc} = 0.35$ eV) than for both crystalline compounds ($E_{dc} = 0.38$ eV). $^7$Li NMR field-gradient measurements of the self-diffusion coefficient $D$ and its activation energy $E_D$ confirm these findings and, furthermore, reveal different Haven ratios. $^7$Li NMR field-cycling relaxometry shows that the lithium ion jumps in ball-milled Li$_{5.5}$PS$_{4.5}$Cl$_{1.5}$ are described by very broad dynamical susceptibilities arising from a temperature-independent Gaussian-like distribution of activation energies $g(E_a)$ with a mean value of $E_m = 0.43$ eV, while the susceptibilities indicated a high-energy cutoff for the crystalline electrolytes. Based on different relations between the activation energies for the conductivity, diffusivity and jumps, we discuss that the shape and exploration of the energy landscapes of ball-milled and crystalline Li$_{5.5}$PS$_{4.5}$Cl$_{1.5}$ samples strongly differ. Moreover, significant differences in the preexponential factor of the *dc* conductivity, the Haven ratio and the single-particle correlation factor point to distinct types of anion lattice disorder of the ball-milled disordered and heat-treated crystalline samples.

**Keywords**

Solid electrolyte, lithium argyrodite, energy landscape, lithium ion dynamics, electrochemical impedance spectroscopy, $^7$Li nuclear magnetic resonance, field-cycling relaxometry, field-gradient diffusometry

## 1. Introduction

In the quest for solid electrolytes with high lithium ionic conductivity, materials with different degrees of structural disorder were at the center of interest [1–6]. Various studies dealt with crystalline materials. For example, there were intense research efforts to improve the charge transport characteristics in crystals by introducing disorder in the mobile ion or immobile ion sublattices [7], leading to compounds with high mobility of lithium ions [8,9]. Other researches focused on amorphous materials with favorable lithium ion transport properties [10]. In the latter approaches, it was shown that the ion mobility can be enhanced by changing the glass composition, e.g., by mixing two network former species [11] or by doping the glasses with lithium halides [12]. Finally, significant work investigated glass ceramics and showed that suitable heat treatment allows one to achieve fast lithium ion dynamics [7,8,13–21]. Therefore, for a development of solid electrolytes with further improved properties, e.g., for use in all-solid-state lithium ion batteries, it is important to understand the similarities and discrepancies of the mechanisms for the jumps and transport of the lithium ion in these materials with strongly different structures. In particular, it is necessary to determine the respective energy landscapes and their role for the lithium ion dynamics on various time and length scales.

For this purpose, lithium argyrodites $Li_6PS_5X$ (X = Cl, Br, I) are interesting materials [14]. They can easily be synthesized in disordered or crystalline forms from cheap starting materials via ball balling or solid-state reaction, possibly followed by appropriate thermal treatment [7]. Moreover, they show high electrochemical and thermal stability and their lithium ion conductivities reach up to several mScm$^{-1}$ at room temperature [15–17], depending on the type of the halide and the synthesis method [18–20]. The high ion conductivities of $Li_6PS_5Cl$ and $Li_6PS_5Br$, which are accompanied by complex motional mechanisms, were rationalized by a high fraction of vacancies in the lithium sublattice and a disordered distribution of the sulfide and halide ions over the anion sites [21–24]. In view of these results, lithium-deficient and halide-rich compositions $Li_{6-x}PS_{5-x}Cl_{1+x}$ were developed to further enhance the lithium ion conductivity [25,26]. For example, $Li_{5.5}PS_{4.5}Cl_{1.5}$ exhibited an ionic conductivity of 9.4 mScm$^{-1}$ at 298 K as a cold pressed pellet, which could be further enhanced to 12 mScm$^{-1}$ by heat treatment [25]. Other promising systems are mixed-halide argyrodites, in particular, crystalline $Li_{6-x}PS_{5-x}ClBr_x$ showing an ionic conductivity of 24 mScm$^{-1}$ for $x = 0.7$ at 298 K [27]. In addition to these crystalline lithium argyrodites with various degrees of disorder in the crystalline lattice, ball-milled disordered $Li_{6-x}PS_{5-x}Cl_{1+x}$ specimen with high lithium ion conductivities were reported [16,28,29]. Hence, lithium argyrodites are ideally suited for a case study on the effects of structural disorder on the energy landscapes and motional mechanisms involved in lithium ion motion and transport in solid electrolytes.

In single-ion conducting materials, complex ion dynamics typically manifest themselves in subdiffusion on short time scales and cooperative diffusion on long time scales. At short times and high frequencies, correlated forward-backward jumps of the ions result in a sublinear increase of their mean square displacement $\langle r^2(t) \rangle$ and, correspondingly in a high-frequency dispersion in the ionic conductivity spectra, explicitly, in a power-law increase of the real part of the ionic conductivity, $\sigma'(\nu)$ [30–34]. On long time scales and low frequencies, ion transport is diffusive, i.e., $\langle r^2(t) \rangle \propto t$, leading to a *dc* conductivity plateau in the conductivity spectra, $\sigma'(\nu) = \sigma_{dc}$. If the diffusive transport occurs in a cooperative manner with distinct ions moving preferentially into the same direction, the *dc* conductivity will be larger than expected from corresponding tracer diffusivities based on the Nernst-Einstein relation and, hence, the Haven ratio $H_R$ will be smaller than unity [35–38]. In addition to electrochemical impedance spectroscopy (EIS) studies of $\sigma(\nu)$, other methods indicated complex mechanisms for ion motion in solid electrolytes. For example, computer simulations revealed correlated forward-backward jumps as well as cooperative ion movements [39–46]. Furthermore, charge attachment induced transport (CAIT) and nuclear magnetic resonance (NMR) experiments revealed strong dynamical heterogeneity [47–52]. In particular, NMR multi-time correlation functions showed that local ion jumps in solid electrolytes are often governed by broad rate distributions. Although these dynamical heterogeneities and the corresponding nonexponentiality of jump correlation functions were found to be particularly prominent for glasses [49,50], they were also observed for various crystals [51,52]. Recently, a combination of EIS and NMR proved to be very useful for studies of solid electrolytes [53,54].

$^7$Li NMR enables valuable insights into short-range and long-range lithium ion dynamics in solid electrolytes [55–58], including lithium argyrodites [14,25,26,59–62]. For example, $^7$Li NMR studies observing the spin-lattice relaxation (SLR) time $T_1$ across $Li_6PS_5X$ compounds not only revealed the relevance of cation and anion disorder for the short-range lithium ion dynamics, but also showed that a complex jump mechanism caused significant deviations from exponential correlation functions and corresponding Lorentzian spectral densities [14,60–62]. Therefore, assuming these functionalities would cause erroneous results, in particular, incorrect activation energies

$E_a$. This problem of conventional $^7$Li SLR analyses can be overcome by using $^7$Li field-cycling relaxometry (FCR), which allows one to measure the $^7$Li SLR time $T_1$ over a broad range of Larmor frequencies $\omega_L$ and, in this way, to map out the spectral density $J_2(\omega_L)$ of the lithium ion jumps [53,63–65]. In addition, $^7$Li NMR field gradient measurements provide access to long-range lithium ion diffusion [66–68]. For $Li_{6-x}PS_{5-x}Cl_{1+x}$ electrolytes, such studies yielded self-diffusion coefficients $D$ as a function of the halide content [25].

In recent work [59], we combined EIS studies with $^7$Li NMR diffusometry and relaxometry to investigate the dynamics and transport of the lithium ions in two crystalline lithium-deficient and halide-rich $Li_{5.5}PS_{4.5}Cl_{1.5}$ argyrodites, which showed different order in the anion sublattice as a result of diverse heat treatments. In EIS, we analyzed the ionic conductivity $\sigma'(\nu)$ over broad frequency and temperature ranges [59]. Moreover, we performed $^7$Li NMR diffusometry in a static field gradient (SFG) to measure the self-diffusion coefficients $D$ of the lithium ions. In doing so, we exploited the fact that using static rather than pulsed magnetic field gradients allowed us to apply stronger gradients and, in this way, to measure lithium ion diffusivities in broader ranges of time and length scales. In $^7$Li FCR experiments, we determined the spectral densities $J_2(\omega_L)$ and corresponding dynamical susceptibilities $\chi''_{NMR}(\omega_L)$ of the lithium ion jumps. For both studied crystalline $Li_{5.5}PS_{4.5}Cl_{1.5}$ argyrodites [59], we found that the conductivity and diffusivity of the lithium ions are related in a straightforward manner with their local jumps. In particular, these dynamical processes were described by similar activation energies.

Here, we use EIS in combination with $^7$Li SFG and FCR studies to compare the energetic and mechanistic characteristics of lithium ion transport and dynamics in a highly disordered $Li_{5.5}PS_{4.5}Cl_{1.5}$ sample, which is prepared by ball milling, with the previously determined ones for the crystalline $Li_{5.5}PS_{4.5}Cl_{1.5}$ compounds [59]. Measurements of $\sigma'(\nu)$ allow us to compare the *dc* conductivities $\sigma_{dc}$ of ball-milled disordered and crystalline $Li_{5.5}PS_{4.5}Cl_{1.5}$ and, by detailed analysis of the high-frequency dispersion, to obtain insight into the respective relevance of subdiffusion for the ion transport. Furthermore, we perform $^7$Li SFG measurements to determine the self-diffusion coefficients $D$ of the lithium ions. Relating the *dc* conductivities $\sigma_{dc}$ with the self-diffusion coefficients $D$ informs about the Haven ratio $H_R$ and, thus, the relevance of correlated lithium ion motion. Finally, we show that the spectral densities $J_2(\omega_L)$ and dynamical susceptibilities $\chi''_{NMR}(\omega_L)$ of the lithium ion jumps from $^7$Li FCR provide straightforward access to the distributions of correlation times $G(\log \tau)$ and, assuming thermally activated motion, the distributions of activation energies $g(E_a)$. Although the thus determined activation energies, in general, depend not only on the energies of the ion sites and of the barriers in between them but also on the occupancy and, hence, the availability of the sites, the information about $g(E_a)$ still allows us to contrast the energy landscapes governing the local ion jumps in samples featuring the same composition but different structures. Last but not least, the activation energies of $D$ and $\sigma_{dc}$, $E_D$ and $E_{dc}$, and, thus of long-range dynamics, are compared with distributions of activation energies $g(E_a)$ for the local jumps to obtain information about the exploration of the energy landscape in disordered and ordered electrolytes.

## 2. NMR background

The $^7$Li nuclei ($I = 3/2$) of the studied samples are subject to the Zeeman interaction, which determines the Larmor frequency $\omega_L = \gamma B_0$, where $\gamma$ denotes the gyromagnetic ratio of the nuclei and $B_0$ is the applied static magnetic field, and the quadrupolar interaction, which depends on the local electric field gradient and, hence, differs at various lithium sites. In $^7$Li FCR, we utilize the relation between the $^7$Li SLR time $T_1$ and the spectral density $J_2(\omega)$ describing the fluctuations of the quadrupolar interactions due to jumps of the lithium ions [55]:

$$\frac{1}{T_1} = C_Q^2 [J_2(\omega_L) + 4 J_2(2\omega_L)] \qquad (1)$$

Here, the coupling constant $C_Q$ is proportional to the strength of the quadrupolar interaction. Thus, the $^7$Li SLR rate $1/T_1$ essentially probes the value of $J_2(\omega)$ at the Larmor frequency $\omega_L$. $^7$Li FCR allows us to measure $T_1$ in broad frequency and temperature ranges [53,63–65]. The resulting data sets $T_1(\omega_L; T)$ can be analyzed in two ways. First, it is possible to study the temperature dependence for several but fixed Larmor frequencies $\omega_L$ so that the respective $T_1(T)$ minima yield correlation times $\tau(T)$ from the relation $\omega_L \tau = 0.616$. Second, the frequency dependence of $1/T_1$ and, thus, of $J_2$ is available for various temperatures $T$. Although the spectral density contains the relevant information, the frequency-dependent analysis profits from switching to a susceptibility representation by considering Eq. (1) and the fluctuation-dissipation theorem, which links spectral densities with the imaginary part of corresponding dynamical susceptibilities, $\chi''(\omega_L) = \omega_L J(\omega_L)$ [53,63–65,69,70]. Specifically, it proved to be beneficial to define the NMR susceptibility

$$\chi''_{\text{NMR}}(\omega_\text{L}) \equiv \frac{\omega_\text{L}}{T_1(\omega_\text{L})} \qquad (2)$$

This representation of FCR data, which often produces a susceptibility peak, enables an interpretation in the analogy with results from electrical and mechanical relaxation studies. In particular, the peak position yields a characteristic correlation time and the peak shape informs about possible dynamical correlations and heterogeneities. In our previous $^7$Li FCR study of crystalline $\text{Li}_{5.5}\text{PS}_{4.5}\text{Cl}_{1.5}$ samples [59], the susceptibility peaks had asymmetric shapes, which could be described by the Havriliak-Negami (HN) function

$$\chi_{\text{HN}}(\omega_\text{L}) = \chi_\infty + \frac{\chi_0 - \chi_\infty}{[1+(i\omega_\text{L}\tau_{\text{HN}})^{\alpha_{\text{HN}}}]^{\beta_{\text{HN}}}} \qquad (3)$$

Here, $\chi_0$ and $\chi_\infty$ are the low-frequency and high-frequency limits, respectively, and $\tau_{\text{HN}}$ is the time constant. Furthermore, $\alpha_{\text{HN}}$ and $\beta_{\text{HN}}$ are shape parameters ($0 < \alpha_{\text{HN}}, \beta_{\text{HN}} \leq 1$), which characterize the slopes of the low-frequency ($+\alpha_{\text{HN}}$) and high-frequency ($-\alpha_{\text{HN}}\beta_{\text{HN}}$) flanks of HN peaks in a double logarithmic representation. For the crystalline $\text{Li}_{5.5}\text{PS}_{4.5}\text{Cl}_{1.5}$ samples [59], we observed an approximate Cole-Davidson (CD) shape, which is obtained from the HN function for $\alpha_{\text{HN}} = 1$ and $\beta_{\text{HN}} < 1$. In the present $^7$Li FCR experiments on ball-milled disordered $\text{Li}_{5.5}\text{PS}_{4.5}\text{Cl}_{1.5}$, we observe nearly symmetric susceptibility peaks and show that the experimental results can be described in a broad range on the basis of a temperature-independent Gaussian distribution of activation energies

$$g_{\text{GS}}(E_\text{a}) = \frac{1}{\sqrt{2\pi}\sigma_\text{E}} \exp\left[-\frac{(E_\text{a} - E_\text{m})^2}{2\sigma_\text{E}^2}\right] \qquad (4)$$

with mean energy $E_\text{m}$ and standard deviation $\sigma_\text{E}$. Specifically, in agreement with previous applications to disordered solid electrolytes [64,65], we use this Gaussian energy distribution $g_{\text{GS}}(E_\text{a})$ together with the Arrhenius law, $\tau = \tau_0 \exp\left(\frac{E_\text{a}}{k_\text{B}T}\right)$, to calculate the corresponding logarithmic Gaussian distribution of correlation times $G_{\text{GS}}(\log \tau)$ at a given temperature and obtain the NMR susceptibility according to:

$$\chi''_{\text{GS}}(\omega_\text{L}) = \int_{-\infty}^{+\infty} [\chi''_\text{D}(\omega_\text{L}\tau) + \chi''_\text{D}(2\omega_\text{L}\tau)] G_{\text{GS}}(\log \tau) \, \text{d}\log \tau \qquad (5)$$

Here, $\chi''_\text{D}(\omega_\text{L}\tau) = \omega_\text{L}\tau/[1 + (\omega_\text{L}\tau)^2]$ denotes the Debye susceptibility for a particular correlation time $\tau$ from the distribution.

In $^7$Li SFG diffusometry, we apply a magnetic field with a static gradient $g$ along the $z$ axis, $B(z) = B_0 + gz$, so that the $^7$Li Larmor frequencies depend on the positions of the lithium ions, $\omega_\text{L}(z) = \gamma B(z)$. Hence, lithium ion diffusion leads to a time dependence of $\omega_\text{L}$, which can be probed when utilizing the stimulated-echo sequence, $90° - t_\text{e} - 90° - t_\text{m} - 90° - t_\text{e}$ to correlate the respective frequencies during two evolution times $t_\text{e}$ separated by a variable mixing time $t_\text{m}$. Assuming free diffusion, the self-diffusion coefficient $D$ of the lithium ions can be obtained from the decay of the observed intensity $S$ of the produced stimulated echo, explicitly, [71]

$$S(t_\text{m}, t_\text{e}) \propto \exp[-Dq^2 t] \qquad (6)$$

Here, $t = t_\text{m} + \frac{2}{3}t_\text{e}$ is the diffusion time and $q = g\gamma t_\text{e}$ can be considered as a generalized scattering vector, the inverse of which determines the length scale of the diffusion measurement. For the used values of the evolution time $t_\text{e}$ and the field gradient $g$, we probe diffusion on length scales in the range of ca. $0.1 - 1$ μm.

## 3. Experimental

### 3.1. Sample preparation

The $\text{Li}_{5.5}\text{PS}_{4.5}\text{Cl}_{1.5}$ sample was prepared by using a high energy planetary ball mill (Pulverisette 7, Fritsch, Idar-Oberstein, Germany). For a 2 g batch, the starting materials were filled into a 20 mL $\text{ZrO}_2$ pot with 10 $\text{ZrO}_2$ balls (∅ = 10 mm) in a stoichiometric mixture of $\text{Li}_2\text{S}$ (99.9 %, Alfa Aesar, Karlsruhe, Germany), $\text{P}_2\text{S}_5$ (for synthesis, Sigma Aldrich, Taufkirchen, Germany), and LiCl (≥99.98%, Sigma-Aldrich, Taufkirchen, Germany). This was done under argon atmosphere in a glovebox (UniLab, MBraun, Garching, Germany; $x_{\text{H}_2\text{O}} < 1$ ppm and $x_{\text{O}_2} < 1$ ppm). The pot was closed air-tight, removed from the glovebox and integrated into the ball mill. Milling was carried out with a rotation speed of 850 rpm for 99 cycles (5 min milling, 15 min rest). Afterwards the pot was transferred into the glovebox and the product was grinded in an agate mortar to obtain $\text{Li}_{5.5}\text{PS}_{4.5}\text{Cl}_{1.5}$. We will refer to this powder as ball-milled (BM) $\text{Li}_{5.5}\text{PS}_{4.5}\text{Cl}_{1.5}$.

The syntheses and characterizations of crystalline samples (2 g batch) was described in a previous work [59]. Briefly, pristine (PR) $Li_{5.5}PS_{4.5}Cl_{1.5}$ was synthesized by solid-state synthesis in a silica ampoule. Further, heat treated (HT) $Li_{5.5}PS_{4.5}Cl_{1.5}$ was obtained by a subsequent heat treatment at 823 K for 10 min.

**3.2. X-ray diffraction measurements**

The X-ray diffraction (XRD) measurements were executed using a powder diffractometer STOE STADI MP (STOE, Darmstadt, Germany) and Cu-Kα radiation in a Debye-Scherrer geometry. The BM $Li_{5.5}PS_{4.5}Cl_{1.5}$ powder was filled into mark tubes (Hilgenberg, Malsfeld, Germany) under argon and closed air-tight by means of a wax.

The XRD pattern of BM $Li_{5.5}PS_{4.5}Cl_{1.5}$ is shown in the supporting material (Fig. S1). The pattern exhibits broad Bragg peaks as well as an amorphous background, suggesting the existence of nanocrystallites (average size 20 – 30 nm) in an amorphous matrix. Furthermore, a small amount of a γ-$Li_3PS_4$ impurity phase [29] is detected.

**3.3. Electrochemical impedance spectroscopy**

EIS measurements were carried out by pressing the BM $Li_{5.5}PS_{4.5}Cl_{1.5}$ powder into pellets with a diameter of 3 mm for 10 min using a fabrication pressure of 84.9 MPa applied by a hydraulic press (P/O/Weber, Remshalden, Germany) with polished stainless steel extrusion dies. Afterwards, the thickness of the BM $Li_{5.5}PS_{4.5}Cl_{1.5}$ pellet was determined with a micrometer caliper (Mitutoyo, Neuss, Germany). A good contact during the measurement was ensured by sputtering gold electrodes onto both faces of the pellet using a sputter coater (108auto, Cressington, Watford, England). Then, the pellet was mounted into a home-built airtight two-electrode cell. The impedance measurements were carried out using an Alpha-AK impedance analyzer (Novocontrol, Montabaur, Germany) with an applied voltage of 10 $mV_{RMS}$ in a frequency range of 1 MHz to 0.1 Hz. The temperature was varied in a range from 153 K to 293 K in 20 K steps using the Novocontrol Quatro Cryosystem with a maximum temperature deviation of ±0.1 K. The impedance spectra were fitted with the software RelaxIS (RHD instruments, Darmstadt, Germany). After the EIS measurements, we removed the gold electrodes from the pellet, filled the pellet into an NMR tube, which was finally sealed under vacuum. Thus, the EIS and NMR measurements were carried out on the identical sample.

**3.3. NMR measurements**

Two home-built NMR setups were employed for the $^7$Li SLR measurements: (i) We used a superconducting magnet operating at a fixed Larmor frequency of $\omega_L = 2\pi \cdot 62.9$ MHz and applied the saturation-recovery sequence. (ii) We utilized a field-cycling relaxometer, which features a switchable electromagnet to observe SLR at adjustable magnetic fields $B_0$ and, thus, Larmor frequencies $\omega_L$ [72]. For the $^7$Li SFG measurements, we employed a specially designed magnet comprising two superconducting coils in an anti-Helmholtz arrangement to produce a magnetic field with strong gradients [73]. The present experiments were performed at a sample position where the field strength and field gradient amounted to $B_0 = 3.8$ T and $g = 144$ Tm$^{-1}$, respectively. To determine self-diffusion coefficients $D$, we recorded stimulated-echo decays $S(t_m)$ for several values of the evolution time $t_e$ and, thus, of $q$, and fitted the results globally with Eq. (4), supplemented by a predetermined factor describing additional SLR damping. In neither of the setups and measurements, the 90° pulse lengths exceeded 2 μs. The temperature was controlled within ±0.5 K by a nitrogen gas flow in the FCR setup and a liquid nitrogen cryostat in the superconducting magnets.

**4. Results**

**4.1. Electrochemical impedance spectroscopy**

The Nyquist plot of the impedance of BM $Li_{5.5}PS_{4.5}Cl_{1.5}$ at a temperature of 173 K is shown in Fig. S2. The experimental data were fitted with the equivalent circuit shown in the inset, and the capacitance of the semicircle was calculated via the Brug formula [74] as shown in [59]. A capacitance of $C_{bulk} = 5 \cdot 10^{-11}$ Fcm$^{-2}$ for the BM sample confirms that the semicircle is caused by the bulk electrical properties of the sample. The *dc* ionic conductivity was calculated from the equation $\sigma_{dc} = (1/R_{bulk}) \cdot (d/A)$, where $d$ and $A$ denote the sample thickness and the electrode area, respectively. In Fig. 1, the *dc* conductivities of BM, PR and HT $Li_{5.5}PS_{4.5}Cl_{1.5}$ are compared in an Arrhenius plot. For all three samples, the temperature dependence of $\sigma_{dc}$ is well described by an Arrhenius law.

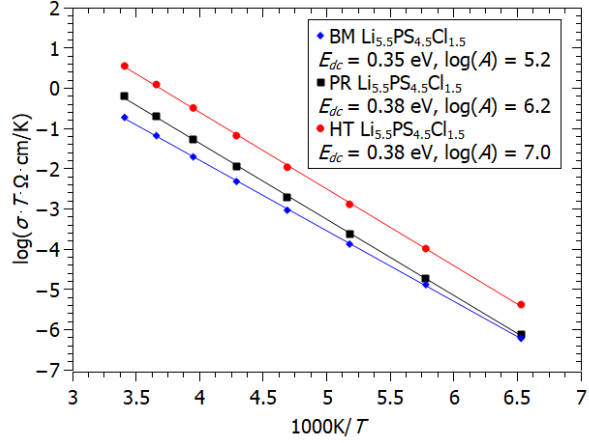

Fig. 1: Arrhenius plot of the *dc* conductivity of BM Li$_{5.5}$PS$_{4.5}$Cl$_{1.5}$ (blue), PR Li$_{5.5}$PS$_{4.5}$Cl$_{1.5}$ (black) and HT Li$_{5.5}$PS$_{4.5}$Cl$_{1.5}$ (red). The solid lines are fits with the Arrhenius equation $\log(\sigma T) = \log(A) - E_{dc}/(k_B T \cdot \ln(10))$ with log denoting the decadic logarithm.

The room-temperature ionic conductivity (at 298 K) of the BM sample is $\sigma_{dc} = 0.76$ mScm$^{-1}$ and is thus lower than $\sigma_{dc} = 2.48$ mScm$^{-1}$ obtained for the PR sample and $\sigma_{dc} = 14.9$ mScm$^{-1}$ obtained for the HT sample, respectively, in previous work [59]. The activation energy of the samples was determined by a fit with the Arrhenius equation $\log(\sigma T) = \log(A) - E_{dc}/(k_B T \cdot \ln(10))$. For BM Li$_{5.5}$PS$_{4.5}$Cl$_{1.5}$, we obtain a value of $E_{dc} = 0.35$ eV as compared to a somewhat higher activation energy of $E_{dc} = 0.38$ eV for PR and HT Li$_{5.5}$PS$_{4.5}$Cl$_{1.5}$ [59]. The decadic logarithm of the preexponential factor is $\log(A) = 5.2$ for the BM sample, compared to $\log(A) = 6.2$ for the PR sample and $\log(A) = 7.0$ for the HT sample. It is important to note that the differences in the room-temperature *dc* conductivity of the samples are mainly caused by differences in $\log(A)$.

In Fig. 2, we show a log-log Summerfield plot of the frequency-dependent ionic conductivity. In this plot, the real part of the complex conductivity $\sigma'$ is normalized to the *dc* conductivity $\sigma_{dc}$, and the frequency $\nu$ is rescaled by the *dc* conductivity multiplied by the temperature. At 173 K, the spectra of all sample exhibit a *dc* plateau at low frequencies and a dispersive regime at high frequencies. We define the crossover frequency $\nu^*$ between the *dc* plateau regime and the dispersive regime as $\sigma'(\nu^*) = 2\sigma_{dc}$ ($\log(\sigma'(\nu^*)/\sigma_{dc}) = 0.3$). The comparison of the results in Fig. 2 reveals that the BM and the PR sample exhibit similar values for the scaled crossover frequency $\nu^*/(\sigma_{dc} \cdot T)$, while the Summerfield plot of the HT sample is characterized by a much higher value of $\nu^*/(\sigma_{dc} \cdot T)$.

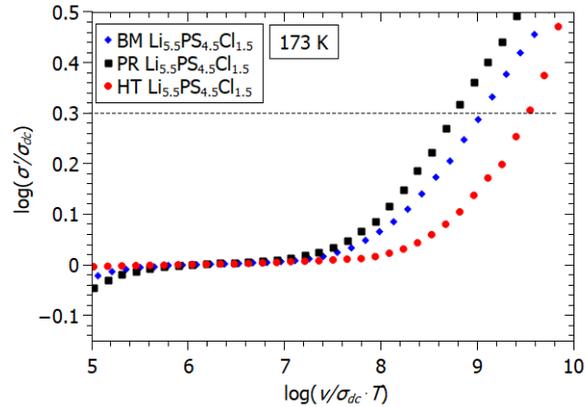

Fig. 2: Summerfield plot of the frequency-dependent conductivity of BM (blue), PR (black) and HT (red) Li$_{5.5}$PS$_{4.5}$Cl$_{1.5}$ at 173 K. The crossover frequency $\nu^*$ is defined by $\log(\sigma'(\nu^*)/\sigma_{dc}) = 0.3$, as indicated by the dashed line.

### 4.2. $^7$Li NMR diffusometry

We move on to $^7$Li NMR studies of lithium ion dynamics in BM Li$_{5.5}$PS$_{4.5}$Cl$_{1.5}$. First, we use $^7$Li SFG NMR to determine the self-diffusion coefficients $D$. For this purpose, we globally fit stimulated-echo decays $S(t_m)$ for various evolution times $t_e$ with Eq. (5). In Fig. 3, the self-diffusion coefficients $D$ resulting from such SFG analyses at various temperatures are compared with those of the previously studied crystalline PR and HT Li$_{5.5}$PS$_{4.5}$Cl$_{1.5}$ samples [59]. We see that, at the studied temperatures, the self-diffusion coefficients $D$ of BM

$Li_{5.5}PS_{4.5}Cl_{1.5}$ are smaller than those of the other samples, consistent with the above findings for the *dc* conductivity $\sigma_{dc}$. At the same time, the activation energy is lower for the disordered material ($E_D = 0.29$ eV) than for both crystalline materials ($E_D = 0.34$ eV) [59]. This observation suggests that lithium ion diffusion becomes faster in BM $Li_{5.5}PS_{4.5}Cl_{1.5}$ than in PR and HT $Li_{5.5}PS_{4.5}Cl_{1.5}$ at lower than the studied temperatures. Comparison with the above EIS results reveals that the activation energies $E_D$ of the diffusivity $D$ are somewhat smaller than those of the *dc* conductivity, $E_{dc}$, but both methods yield the same order across the studied electrolytes.

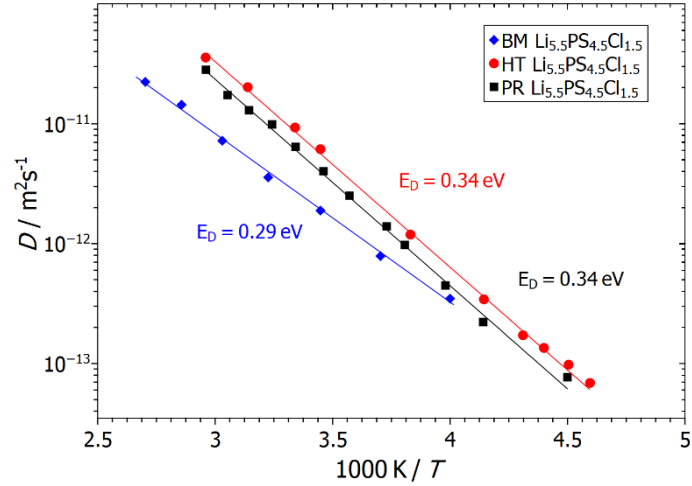

Fig. 3: Self-diffusion coefficients $D$ of the lithium ions in BM (blue), PR (black), and (red) HT $Li_{5.5}PS_{4.5}Cl_{1.5}$. The results for both latter samples were obtained in previous work [59]. The lines are fits with Arrhenius laws, yielding activation energies of $E_D = 0.29$ eV for BM $Li_{5.5}PS_{4.5}Cl_{1.5}$ and $E_D = 0.34$ eV for PR and HT $Li_{5.5}PS_{4.5}Cl_{1.5}$ [59].

### 4.3. $^7$Li NMR relaxometry

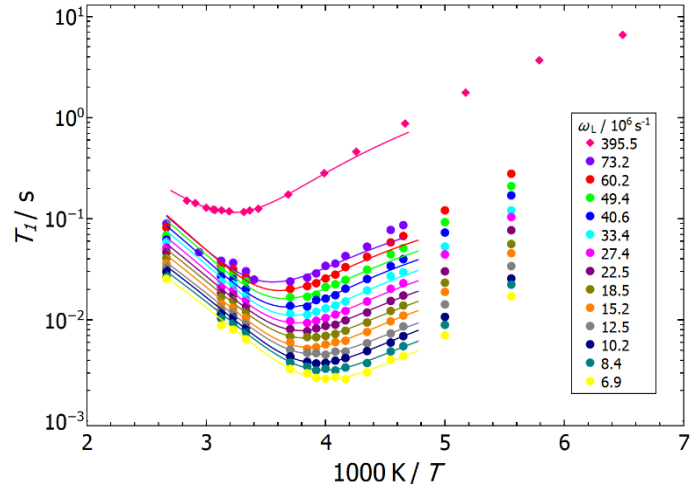

Fig. 4: Temperature-dependent $^7$Li SLR times $T_1$ of BM $Li_{5.5}PS_{4.5}Cl_{1.5}$ for the indicated Larmor frequencies $\omega_L$. The data at $\omega_L = 395.5 \cdot 10^6$ s$^{-1}$ were measured in a superconducting magnet, while those at lower Larmor frequencies were obtained with the field-cycling setup. The lines are parabolic fits, which serve to determine the $T_1(T)$ minima.

Having determined the long-range dynamics of the lithium ions, we next employ $^7$Li FCR to investigate their next-neighbor jumps. First, we exploit the fact that $^7$Li FCR yields $T_1(T)$ for various values of $\omega_L$ so that the respective minima can be used to determine correlation times. Fig. 4 displays exemplary $T_1(T)$ minima for BM $Li_{5.5}PS_{4.5}Cl_{1.5}$ and various Larmor frequencies. Specifically, data from FCR studies at $\omega_L < 100 \cdot 10^6$ s$^{-1}$ are complemented by $T_1$ values from conventional SLR measurements at $\omega_L = 395 \cdot 10^6$ s$^{-1}$. The $T_1(T)$ minima shift to lower temperatures when the Larmor frequency is decreased, reflecting the slowdown of the lithium ion jumps upon cooling. For further analysis, we obtain the positions of the respective minima from parabolic fits and use the condition $\omega_L \tau = 0.616$ to assign correlation times $\tau$ to these temperatures.

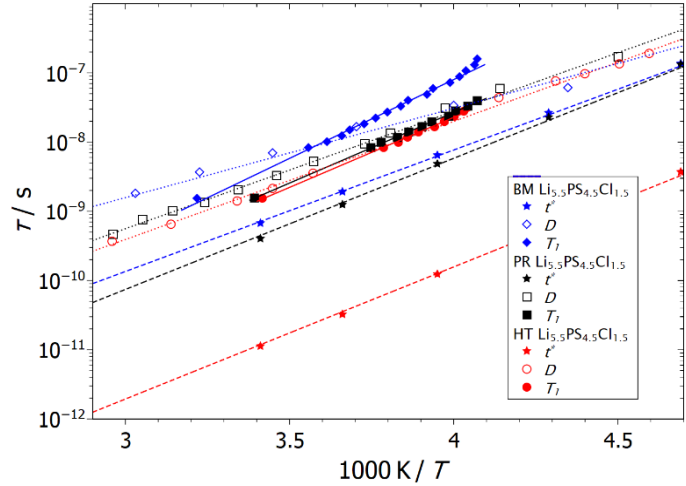

Fig. 5: Temperature-dependent correlation times $\tau$ of lithium ion dynamics in the BM (blue), PR (black) [59] and HT (red) [59] $Li_{5.5}PS_{4.5}Cl_{1.5}$ electrolytes. Solid symbols: correlation times determined from the $^7Li\ T_1(T)$ minima for various Larmor frequencies $\omega_L$. Solid lines: Arrhenius fits of the $^7Li\ T_1(T)$ results, yielding activation energies of $E_{T1} = 0.45$ eV for BM $Li_{5.5}PS_{4.5}Cl_{1.5}$ and $E_{T1} = 0.40$ eV for PR and HT $Li_{5.5}PS_{4.5}Cl_{1.5}$. Open symbols: correlation times calculated from the self-diffusion coefficients $D$, see Fig. 3, using $\tau = a^2/(6D)$ with $a = 2.81$ Å [63]. Dotted lines: Arrhenius fits of the calculated diffusion correlation times. The activation energies are specified in Fig. 3. Stars: correlation times obtained from the crossover frequency in the conductivity spectra $\sigma'(\nu)$ according to $t^* = 1/(2\pi\nu^*)$. Dashed lines: Arrhenius fits of $t^*$, yielding activation energies of 0.35 eV for the BM sample and of 0.38 eV for the PR and HT samples.

In Fig. 5, we see that the correlation times $\tau$ from the $^7Li\ T_1(T)$ analysis are longer for BM $Li_{5.5}PS_{4.5}Cl_{1.5}$ than for PR and HT $Li_{5.5}PS_{4.5}Cl_{1.5}$. Thus, the former sample not only shows slower lithium ion transport than the crystalline argyrodites but also slower lithium ion jumps. However, the activation energies obtained from different observables do not evolve consistently across the studied samples. While the conductivity and diffusivity activation energies are smaller for BM $Li_{5.5}PS_{4.5}Cl_{1.5}$ than for PR and HT $Li_{5.5}PS_{4.5}Cl_{1.5}$, see Figs. 1 and 3, the opposite is true for the activation energies of the jump correlation times. Specifically, $E_{T1} = 0.45$ eV results from the present $^7Li\ T_1(T)$ study for the BM sample, whereas $E_{T1} = 0.40$ eV was previously obtained from an analogous approach to both crystalline compounds [59]. This discrepancy is also evident when we relate the dynamics on different length scales by assuming that the lithium ions perform a random walk described by a single jump correlation time $\tau$ and jump length $a$ so that the relation $\tau = a^2/(6D)$ holds. In previous work [59], we showed that, for PR and HT $Li_{5.5}PS_{4.5}Cl_{1.5}$, the correlation times measured in the $^7Li\ T_1(T)$ study matched very well with those calculated from the self-diffusion coefficients $D$ when using this relation and identifying the jump length with the inter-cage distance in crystalline $Li_{5.5}PS_{4.5}Cl_{1.5}$, $a = 2.81$ Å [63]. However, this approach fails for BM $Li_{5.5}PS_{4.5}Cl_{1.5}$. Explicitly, we see in Fig. 5 that, unlike for the crystalline argyrodites, the measured correlation times show a notably stronger temperature dependence than those calculated from the diffusivities for the disordered electrolyte. Finally, we compare the data from the $^7Li\ T_1(T)$ analysis with correlation times obtained from the crossover frequency in the conductivity spectra $\sigma'(\nu)$ according to $t^* = 1/(2\pi\nu^*)$ in Fig. 5. The correlation times from the SLR and EIS analyses have the same order across the $Li_{5.5}PS_{4.5}Cl_{1.5}$ samples. However, the $^7Li\ T_1$ correlation times are longer than those characterizing the conductivity crossover frequency for all studied $Li_{5.5}PS_{4.5}Cl_{1.5}$ samples. In particular, for HT $Li_{5.5}PS_{4.5}Cl_{1.5}$, the previous $^7Li\ T_1(T)$ analysis observed ion jumps occurring well inside the low-frequency *dc* plateau regime rather than in the high-frequency dispersive regime of $\sigma'(\nu)$. For BM $Li_{5.5}PS_{4.5}Cl_{1.5}$, the difference between the $^7Li\ T_1$ and conductivity correlation times is overall weaker and diminishes when increasing the temperature.

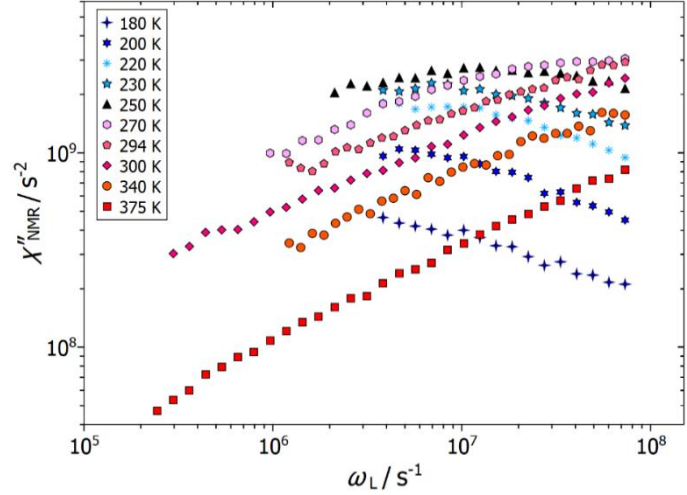

Fig. 6: NMR susceptibility $\chi''_{NMR}(\omega_L)$ of BM $Li_{5.5}PS_{4.5}Cl_{1.5}$ at exemplary temperatures.

Finally, we use $^7$Li FCR to determine the frequency dependence of the SLR time, $T_1(\omega_L)$. In Fig. 6, we show the NMR susceptibility $\chi''_{NMR}(\omega_L) = \omega_L/T_1(\omega_L)$ of BM $Li_{5.5}PS_{4.5}Cl_{1.5}$ obtained from FCR studies at various temperatures. We see a susceptibility peak, which shifts to lower frequencies upon cooling, reflecting the slowdown of the lithium ion jumps. At 250 K, the peak of $\chi''_{NMR}(\omega_L)$ is located at $\omega_L \approx 10^7$ s$^{-1}$, indicating a typical correlation time of $\tau \approx 10^{-7}$ s. At high and low temperatures, we merely observe the low-frequency and high-frequency flanks of the peak, respectively. In the used double logarithmic representation, it is nevertheless evident that the flanks do not exhibit the $\omega_L^{+1}$ and $\omega_L^{-1}$ frequency dependencies, which are characteristic for a Debye peak $\chi''_D(\omega_L)$ associated with an exponential correlation function and a Lorentzian spectral density. Rather, more shallow slopes indicate a significant peak broadening, as expected for heterogeneous lithium ion jumps governed by a broad distribution of correlation times due to an underlying disordered energy landscape.

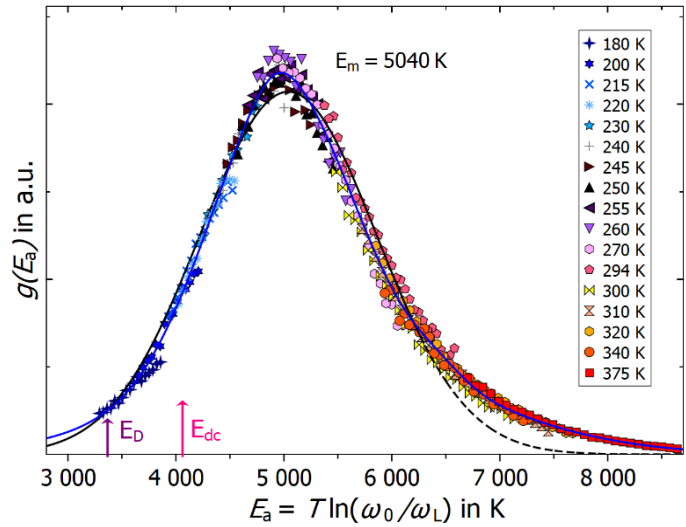

Fig. 7: Distribution of activation energies $g(E_a)$ of BM $Li_{5.5}PS_{4.5}Cl_{1.5}$ obtained by rescaling the $^7$Li NMR susceptibilities $\chi''_{NMR}(\omega_L)$ at the indicated temperatures (180 – 375 K) according to Eq. (7), i.e., by plotting $g(E_a) \sim \chi''_{NMR}(\omega_L)/T$ vs. $E_a = T \ln(\omega_0/\omega_L)$. An attempt frequency of $\omega_0 = 1/\tau_0 = 1.47 \cdot 10^{16}$ s$^{-1}$ was used as obtained from the preexponential factor of the Arrhenius fit of the correlation times from the $T_1$ minima, see Fig. 5. The black line is an interpolation with a Gaussian distribution of activation energies characterized by $E_m = 0.434$ eV (5040 K) and $\sigma_E = 0.083$ eV (960 K). The solid black line indicates the fitted data range, while the dashed black line is an extension of the Gaussian distribution to higher activation energies $E_a$. The arrows indicate the diffusivity ($E_D = 0.29$ eV) and conductivity ($E_{dc} = 0.35$ eV) activation energies, respectively. The blue line is a fit with an arbitrary function, which merely serves to interpolate the data in the whole $E_a$ range.

Therefore, we follow previous $^7$Li FCR studies [64,65] and assume thermally activated jumps governed by a temperature-independent distribution of activation energies $g(E_a)$. In such a situation, the Arrhenius law allows us to obtain the distribution of correlation times $G(\log \tau)$ at a given temperature from the distribution of activation energies $g(E_a)$. Furthermore, $\chi''_{NMR}(\omega_L)$ can be described as a superposition of the susceptibility contributions associated with the various correlation times from the distribution, see Eq. (5). However, in the present case, the observed susceptibility peak is too broad and the experimental frequency range is too narrow to reliably determine the distribution of correlation times $G(\log \tau)$ and, thus, the underlying distribution of activation energies $g(E_a)$ from fits of the experimental data for any of the studied temperatures. This problem can be overcome, when we exploit the fact that, for broad susceptibility peaks, $g(E_a)$ is directly available from plotting $\chi''_{NMR}(\omega_L)$ on rescaled axes, explicitly, from [53,75,76].

$$g\left(E_a = T\ln\left(\frac{\omega_0}{\omega_L}\right)\right) \propto \frac{\chi''_{NMR}(\omega_L)}{T} \qquad (7)$$

Here, $\omega_0 = 1/\tau_0$ is the attempt frequency of the Arrhenius law. It can be obtained from the Arrhenius fit of the FCR results for BM $Li_{5.5}PS_{4.5}Cl_{1.5}$ in Fig. 5, yielding $\tau_0 = 6.8 \cdot 10^{-17}$ s. Thus, this scaling approach does not involve any free parameter and merely relies on the assumptions that the Arrhenius law is valid and the distribution of activation energies is temperature independent. In Fig. 7, we see that the individual susceptibility data for all studied temperatures 180 – 375 K collapse onto a master curve when scaling the axes according to Eq. (7). This collapse involves a wide range of activation energies (~ 3500 – 8500 K) and confirms the validity of our assumption of thermally activated lithium ions jumps governed by a temperature-independent distribution $g(E_a)$. Inspection of the master curve suggests that the distribution of activation energies has a Gaussian shape with some deviations at higher $E_a$ values. Therefore, we perform a Gaussian fit of the data at $E_a \leq 6300$ K. We find that the majority of the lithium ion jumps in BM $Li_{5.5}PS_{4.5}Cl_{1.5}$ is described by a Gaussian distribution of activation energies with a mean value of $E_m = 5040$ K (0.43 eV) and a standard deviation of $\sigma_E = 960$ K (0.08 eV). We expect that the Gaussian distribution reflects the high structural disorder of BM $Li_{5.5}PS_{4.5}Cl_{1.5}$, while the deviations at higher $E_a$ values result from more ordered regions showing up in the XRD data.

## 5. Discussion

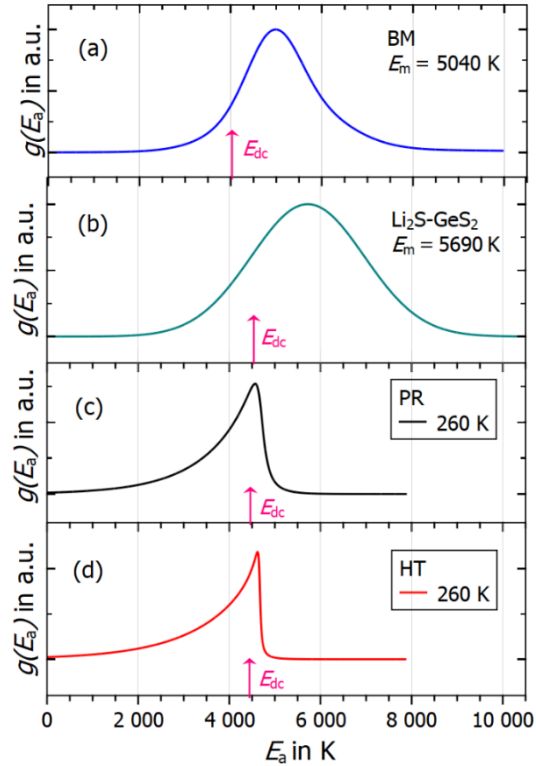

Fig. 8: Distribution of activation energies $g(E_a)$ of (a) BM $Li_{5.5}PS_{4.5}Cl_{1.5}$, (b) $Li_2S$-$GeS_2$ glass, (c) PR $Li_{5.5}PS_{4.5}Cl_{1.5}$, and (d) HT $Li_{5.5}PS_{4.5}Cl_{1.5}$. For the BM sample, the fit, which interpolates the data in the whole $E_a$ range, is shown, see Fig. 7. The data for $Li_2S$-$GeS_2$ glass are a Gaussian distribution [$E_m = 0.49$ eV (5690 K) and $\sigma_E = 0.10$ eV] obtained from a global fit of $^7Li$ FCR susceptibilities at various temperatures in previous work [64]. The data for the crystalline argyrodites were calculated from previously determined HN fit parameters [59], using Eqs. (8) and (9) and a temperature of 260 K.

We proceed by discussing what information about the effects of disorder on the energy landscape and motional mechanism for the lithium ion dynamics and transport in $Li_{5.5}PS_{4.5}Cl_{1.5}$ electrolytes is available from our EIS, SFG, and FCR findings. First, we compare the energy landscapes of the ball-milled and crystalline $Li_{5.5}PS_{4.5}Cl_{1.5}$ electrolytes on the basis of our $^7Li$ FCR results. The present study revealed very broad and essentially symmetric NMR susceptibilities $\chi''_{NMR}(\omega_L)$ for BM $Li_{5.5}PS_{4.5}Cl_{1.5}$. Specifically, a scaling approach showed that the majority of the lithium ion jumps in disordered BM $Li_{5.5}PS_{4.5}Cl_{1.5}$ was described by a Gaussian distribution of activation energies with a mean value of $E_m = 0.43$ eV and a standard deviation of $\sigma_E = 0.08$ eV, see Fig. 8(a). In Fig. 8(b), it can be seen that a similar Gaussian distribution $g(E_a)$ was found in a previous $^7Li$ FCR approach to $Li_2S$-$GeS_2$ glass, for which $E_m = 0.49$ eV and $\sigma_E = 0.10$ eV were reported [64]. By contrast, $^7Li$ FCR previously yielded strongly asymmetric susceptibilities for PR and HT $Li_{5.5}PS_{4.5}Cl_{1.5}$ [59]. In detail, a CD-like shape of the NMR susceptibility peak was indicated by fits with a HN function, yielding approximate $\omega_L^{+1}$ behavior on the low-frequency flank ($\alpha_{HN} \approx 1$) together with a much more shallow high-frequency flank ($\beta_{HN} \approx 0.2$). However, the previous $^7Li$ FCR analysis did not assume a temperature-independent distribution of activation energies for the crystalline $Li_{5.5}PS_{4.5}Cl_{1.5}$ compounds, but it proved to be suitable to suppose frequency-temperature superposition. Nonetheless, it is still possible to calculate $g(E_a)$ for a specific temperature from the HN fits. For this purpose, we first determine the HN distribution of correlation times from the previously obtained fit parameters [59], see Eq. (3), according to [77].

$$g(\ln \tau) = \frac{1}{\pi} \frac{(\tau/\tau_{HN})^{\alpha_{HN}\beta_{HN}} \sin(\beta_{HN}\theta)}{((\tau/\tau_{HN})^{2\alpha_{HN}} + 2(\tau/\tau_{HN})^{\alpha_{HN}} \cos(\pi\alpha_{HN}) + 1)^{\beta_{HN}/2}} \quad (8)$$

$$\theta = \arctan\left(\frac{\sin(\pi\alpha_{HN})}{(\tau/\tau_{HN})^{\alpha_{HN}} + \cos(\pi\alpha_{HN})}\right) + c \quad (9)$$

Here, $c = \pi$ if the argument of the arctangent is negative and $c = 0$ otherwise [77]. Afterward, we use the Arrhenius law to convert the distribution of correlation times into the corresponding distribution of activation energies at a given temperature. Figures 8(c) and 8(d) display $g(E_a)$ calculated in this way for PR and HT $Li_{5.5}PS_{4.5}Cl_{1.5}$ at a temperature of 260 K. A comparison of all results in Fig. 8 reveals that $g(E_a)$ has a strongly different shape in the ball-milled and glassy electrolytes as compared to the crystalline electrolytes. Specifically, the broad Gaussian distributions of activation energies for the ball-milled and glassy electrolytes samples are clearly distinguishable from the highly skewed distributions with a high-energy cutoff for the crystalline argyrodites. In our previous study [59], we argued that the latter cutoff results from the cage-like arrangement of the lithium sites in crystalline argyrodites, which means that an inter-cage jump process of the lithium ions is well defined and has the longest correlation time and the highest activation energy. Therefore, we propose that the absence of a high-energy cutoff for BM $Li_{5.5}PS_{4.5}Cl_{1.5}$ indicates a partial distortion of the cage structure in this disordered material. In passing, we note that minor deviations from a CD-like shape, i.e., from a sharp cutoff, were slightly more prominent for PR than HT $Li_{5.5}PS_{4.5}Cl_{1.5}$ argyrodite because minor impurities with slower lithium ion dynamics and, hence, higher activation energies, which existed in the PR sample, were removed during the additional heat treatment used to produce the HT sample [59].

To gain insight into the relation between short-range and long-range lithium ion dynamics, we next relate the distributions $g(E_a)$ describing the ion jumps with the activation energy of the dc conductivity, $E_{dc}$. In Fig. 8, we observe that, for PR and HT $Li_{5.5}PS_{4.5}Cl_{1.5}$ argyrodite, $E_{dc}$ is located near the high-energy cutoff of $g(E_a)$. This is in agreement with the above argument that, in crystalline argyrodites, well-defined inter-cage jumps have the longest correlation time and the highest activation energy, but are a requisite for long-range charge transport and, hence, yield the relevant energy barrier against dc conductivity. The situation is completely different for BM $Li_{5.5}PS_{4.5}Cl_{1.5}$ and $Li_2S$-$GeS_2$ glass [64]. In these cases, $E_{dc}$ is located in the low-energy wing of the Gaussian distribution $g(E_a)$, i.e., it is substantially smaller than $E_m$. Specifically, integration of $g(E_a)$ up to $E_{dc}$ shows that only 11.4 % and 15.2 % of the lithium ion jumps are governed by activation energies $E_a < E_{dc}$ in BM $Li_{5.5}PS_{4.5}Cl_{1.5}$ and in $Li_2S$-$GeS_2$ glass, respectively. Differences between $E_{dc}$ and $E_m$ are expected in percolation approaches to the ionic conductivity in amorphous electrolytes [78–80]. Specifically, for a random barrier model and simple cubic lattice with a coordination number of $z = 6$, a percolation threshold of about 25 % is expected [81]. This value is somewhat higher but still comparable to the fractions of $g(E_a)$ at $E_a < E_{dc}$ for the disordered electrolytes in Figs. 8, suggesting that percolation arguments are relevant. In particular, we would like to emphasize that (i) the percolation threshold strongly depends on the correlation number [80] and a distribution of $z$ is expected for disordered materials and (ii) the percolation threshold is somewhat different when allowing for a distribution of site energies, which is expected based on results of CAIT experiments for various amorphous electrolytes [47,48].

To continue our discussion about how the distinct distributions of activation energies $g(E_a)$ and structural differences of the anion lattice influence the long-range ion transport, we consider three different transport quantities: (i) The preexponential factor of the *dc* conductivity, $A$, which should be influenced by the number of available transport pathways [81]. As mentioned in the Results section, the differences in the *dc* conductivity of the samples are mainly caused by differences in $\log(A)$. (ii) The Haven ratio $H_R$, which is defined as [38]:

$$H_R \equiv \frac{\lim_{t\to\infty}\frac{d}{dt}\sum_{i=1}^{N}(\Delta\vec{R}_i(t))^2}{\lim_{t\to\infty}\frac{d}{dt}(\sum_{i=1}^{N}\Delta\vec{R}_i(t))^2} = \frac{cF^2 D}{RT\,\sigma_{dc}} \qquad (10)$$

Here, $\Delta\vec{R}_i(t)$ is the displacement vector of ion $i$ after the time $t$. Moreover, $c$ denotes the molar concentration of the ions, while $F$ and $R$ are the Faraday constant and the gas constant, respectively. If distinct ions preferentially move in the same direction, the cross-terms in the denominator will, on average, yield positive contributions, resulting in a Haven ratio $H_R < 1$, as usually found for solid electrolytes [35–38]. (iii) The single-particle correlation factor $f$, which is defined as [38]:

$$f = \frac{\langle r^2(t^*)\rangle}{2a^2} = \frac{\frac{1}{N}\sum_{i=1}^{N}(\Delta\vec{R}_i(t^*))^2}{2a^2} = \frac{6\,\sigma_{dc}\,RT H_R}{cF^2 a^2 \nu^*} \qquad (11)$$

Here, $\langle r^2(t^*)\rangle$ denotes the (single-particle) mean square displacement of the ions at the crossover time $t^*$ and $a$ is again the jump length. Single-particle correlation factors $f \ll 1$ and $f \approx 1$ imply weakly pronounced and strongly pronounced subdiffusive single-particle ion dynamics, respectively. We calculate $f$ using the *dc* conductivity $\sigma_{dc}$ and crossover frequency $\nu^*$ from the conductivity spectrum at 173 K, a jump length of $a = 2.81$ Å [63], and the Haven ratio. In doing so, we assume that $H_R$ has a negligible temperature dependence.

Tab. 1: Logarithm of the preexponential factor of the *dc* conductivity, $\log(A)$, Haven ratio $H_R$ (298 K), and single-particle correlation factor $f$ (173 K) of the lithium ion dynamics in the BM, PR, and HT samples.

|  | $\log(A)$ | $H_R$ | $f$ |
|---|---|---|---|
| BM Li$_{5.5}$PS$_{4.5}$Cl$_{1.5}$ | 5.2 | 0.77 | 0.77 |
| PR Li$_{5.5}$PS$_{4.5}$Cl$_{1.5}$ | 6.2 | 0.57 | 0.95 |
| HT Li$_{5.5}$PS$_{4.5}$Cl$_{1.5}$ | 7.0 | 0.13 | 0.04 |

In our previous study on the crystalline argyrodites [59], we assumed that the structure of the PR sample is characterized by an anion-ordered lattice, while heat treatment leads to an anion-disordered lattice in the HT sample, see Fig. 9. The disordered BM sample should also be characterized by an anion-disordered structure, however, since the three transport quantities listed in Tab. 1 differ considerably for the HT sample and for the BM sample, the anion disorder should be clearly distinct in the two samples. Morgan suggested an anion disorder in crystalline argyrodites, which is characterized by percolating transport pathways for lithium ions along PS$_4^{3-}$ groups [24]. This type of anion disorder should lead to a large number of string-like lithium ion transport pathways resulting in a high preexponential factor of the *dc* conductivity, to highly cooperative lithium ion transport along these string-like pathways resulting in a low Haven ratio, and to weakly pronounced lithium ion subdiffusion on short times scales resulting in a low single-particle correlations factor $f$. In addition, the activation energy of the *dc* conductivity, $E_{dc}$, should be located near the high-energy cutoff of the distribution function of activation energies $g(E_a)$. Considering the values listed in Tab. 1 for the HT sample and reinspecting Fig. 8(d), it seems likely that this type of anion disorder exists in the HT sample.

For the BM sample, on the other hand, the preexponential factor of the *dc* conductivity is much lower than for the HT sample, see Tab. 1, indicating much less lithium ion transport pathways in the former disordered sample. A low number of transport pathways is further corroborated by the finding that $E_{dc}$ is located in the low-energy wing of the Gaussian distribution $g(E_a)$, see Fig. 7. Based on these findings, we suggest that the anion disorder in the disordered BM sample is characterized by a random spatial distribution of PS$_4^{3-}$ groups, S$^{2-}$ anions, and Cl$^-$ anions and that lithium ion transport preferably takes place along pathways with Li-S coordination, see Fig. 9. This leads to a small number of percolating transport pathways and to many dead-end pathways. The high number of dead-end pathways results in a strongly pronounced subdiffusion of the lithium ions on short times scales, explaining why the single-particle correlation factor $f$ is much higher for the BM sample than for the HT sample. In addition, the random distribution of anionic groups allows for some cooperative string-like lithium ion transport. Yet, the fraction of suitable pathways and, hence, the fraction of lithium ions showing cooperative string-like motion in the subdiffusive regime at $t < t^*$ is lower in the BM sample than in the HT sample. This results in a Haven ratio $H_R = 0.77$ for the BM sample, which is considerably higher than $H_R = 0.13$ of the HT sample. However, we emphasize that different groups of lithium ions are involved in such string-like motion in the course of time and, hence, a single self-diffusion coefficient $D$ is established in the diffusive regime at $t > t^*$, as indicated by the observation of exponential $^7$Li SFG $S(t_m)$ decays.

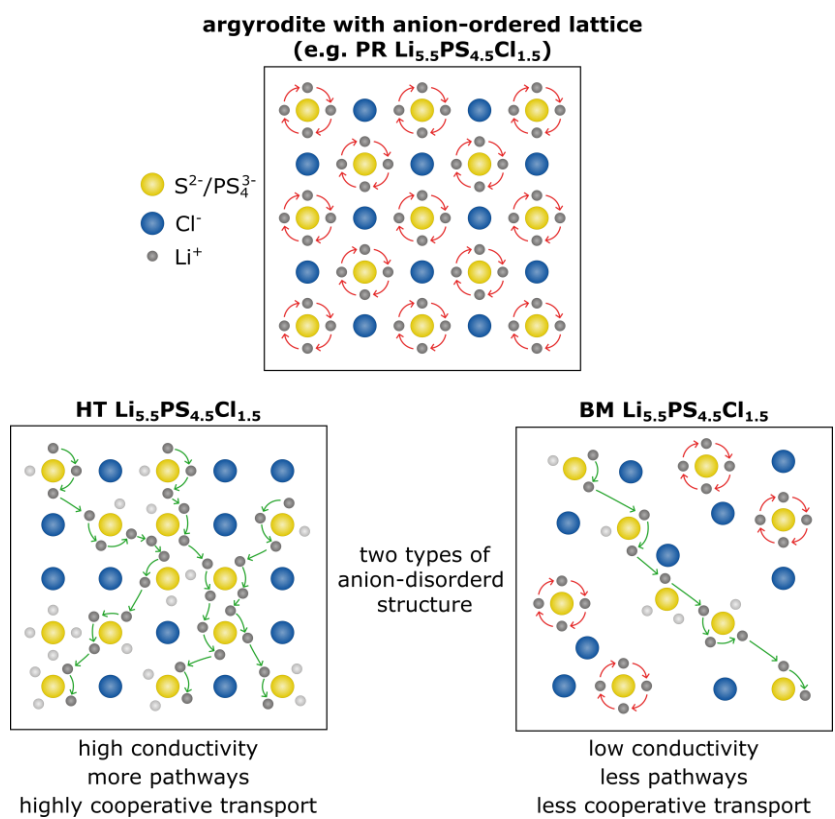

Fig. 9: Schematic illustration of the influence of disorder in the anion lattice on the dynamics and transport of the lithium ions.

## 6. Conclusions

A combination of electrochemical impedance spectroscopy, in particular, detailed analysis of conductivity spectra $\sigma'(\nu)$, with $^7$Li NMR field-gradient diffusometry and field-cycling relaxometry enabled valuable insights into the energy landscapes and motional mechanisms for fast jumps and transport of lithium ions in ball-milled disordered and crystalline $Li_{5.5}PS_{4.5}Cl_{1.5}$ electrolytes. These fast lithium ion conductors showed very rich dynamics, which strongly depended on the respective structures.

The electrochemical impedance data and the field-gradient diffusometry data were used to extract information on the preexponential factor of the *dc* conductivity, on the Haven ratio, and on the single-particle correlation factor. A comparison of these quantities for the BM sample and the HT sample gives strong indication that the disorder of the anion lattice is clearly distinct in these two materials. The lithium ion transport in the HT sample appears to be governed by a large number of percolating transport pathways along $PS_4^{3-}$ groups and $S^{2-}$ anions, leading to highly cooperative long-range lithium ion transport and to weakly pronounced lithium ion subdiffusion on short times scales. In contrast, we suggest that in the BM sample, a random spatial distribution of $PS_4^{3-}$ groups, $S^{2-}$ anions and $Cl^-$ anions results in a lower number of percolating transport pathways and to pronounced subdiffusive ion dynamics. The different numbers of percolating transport pathways in the HT and BM samples result in significantly different Haven ratios $H_R$.

Finally, field-cycling relaxometry revealed that the different microscopic structures manifest themselves in distinct energy landscapes. For crystalline $Li_{5.5}PS_{4.5}Cl_{1.5}$, the cage-like organization of the lithium ion sites leads to a distribution of activation energies with a high-energy cutoff associated with inter-cage jumps, which govern the long-range lithium ion transport. Contrarily, for disordered BM Li$_{5.5}$PS$_{4.5}$Cl$_{1.5}$, the distribution of activation energies does not show such a high-energy cutoff, but rather has a Gaussian-like shape, suggesting that a fraction of the cage-like structure is destroyed. As a result, inter-cage jumps lose their meaning and percolation aspects become relevant for the long-range lithium ion transport, resulting in an activation energy $E_{dc}$, which is substantially smaller than the mean activation energy $E_m$ of the lithium ion jumps, resembling the situation in glassy electrolytes.

**Declaration of competing interest**

The authors declare that they have no known competing financial interests or personal relationships that could have appeared to influence the work reported in this paper.

**Data availability**

Data will be made available on request.

**Acknowledgements**

M. B. and M. V. thank the Deutsche Forschungsgemeinschaft (DFG) for funding in the framework of FOR 5065 (VO 905/13-1). V. M. and B. R. thank the Deutsche Forschungsgemeinschaft (DFG) for funding in the framework of the projects RO 1213/20-1 and RO 1213/14-2.